\documentclass[twocolumn,superscriptaddress,showpacs,aps,prl]{revtex4}
\usepackage{makeidx}
\usepackage{bm}
\usepackage{graphics}
\usepackage{epsfig}

\newcounter{fig}

\begin{document}

\title{ Space-time dispersion of graphene conductivity}
\author{L.A. Falkovsky}
\affiliation{L.D. Landau Institute for Theoretical Physics, Moscow 117334, Russia}
\affiliation{Institute of the High Pressure Physics, Troitsk 142190, Russia}
\author{A.A. Varlamov}
\affiliation{COHERENTIA -INFM, CNR, via del Politecnico 1, I-00133, Rome, Italy }
\pacs{81.05Uw, 78.67.Ch, 78.67.-n}

\begin{abstract}
We present an analytic calculation of the conductivity of pure
graphene as a function of frequency $\omega $, wave-vector $k$,
and temperature for the range where the energies related to all
these parameters are small in comparison with the band parameter
$\gamma =3$ eV. The simple asymptotic expressions are given in
various limiting cases. For instance, the conductivity for
$kv_{0}\ll T\ll \omega $ is equal to $\sigma (\omega
,k)=e^{2}/4\hbar $ and independent of the band structure
parameters $\gamma $ and $v_{0}$. Our results are also used to
explain the known dependence of the graphite conductivity on
temperature and pressure.
\end{abstract}

\maketitle


\section{Introduction}

Recently the properties of graphene attracted special attention (see, for
instance \cite{Novo} and references therein). The matter of fact the wide
variety of carbon materials such as graphite, graphene tubules, and
fullerenes consists of graphene (i.e., a single layer of graphite) sheets
shaped in different manner.

Two-dimensional graphene has a very simple band structure, which can be
obtained with help of the symmetry consideration or in tight-binding
approximation \cite{SDD}. It was shown in Refs. \cite{W,SW} that the energy
bands of graphene are degenerated at the corners of the 2D Brillouin zone $%
K=(0,4\pi /3\sqrt{3}a)$, where $a$=1.44 $\mathring{A}$ is the interatomic
distance (see Fig. 1). This is the Dirac-type spectrum but it is massless
and two-dimensional. It was demonstrated that due to the symmetry arguments
such gapless spectrum with the conic point in the 3D case turns out stable
with respect to the Coulomb interaction \cite{AB70}. One can proof that this
stability remains also for the 2D graphene spectrum with the conic point.
\begin{figure}[h]
\resizebox{.15\textwidth}{!}{\includegraphics{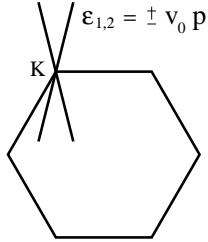}}
\caption{The Brillouin zone of graphene and the electron spectrum in the
vicinity of the $K$ points}
\label{0}
\end{figure}
Such simple band structure can be used in analytic calculations of various
thermodynamic and transport properties of graphene. An example of such
calculations was presented in Ref. \cite{Pe} where the imaginary part of the
dielectric function Im$\varepsilon (\omega )$ of graphite has been
calculated. The gapless band structure of graphene also results in the
unusual behavior of conductivity. Despite the enormous number of paper
devoted to carbon materials, the analytical expression of graphene
conductivity was not yet derived to our best knowledge.

The standard diagrammatic approach to calculation of transport properties of
impure metals is restricted by the fundamental requirement that the mean
free path of carriers $\ell =v_{0}\tau $ must be much larger than the
electron wavelength $\lambda =h/p_{F}$, i.e. $\ell p_{F}\gg 1$. This
condition cannot be evidently satisfied in the case of graphene when the
Fermi surface degenerates to the points. We avoid this difficulty addressing
the problem when the graphene sample is clean enough but temperature is
finite ($T\tau \gg 1$). In this case the temperature appears instead of
Fermi energy and electrons obey Boltzmann statistics.

In the present paper, we calculate analytically the frequency and
wave-vector dependence of the graphene conductivity $\sigma _{ij}(\omega ,k)$
at finite temperature $T$ for the relatively low frequencies $\omega \ll 3$
eV, when only the nearest $\pi $ bands can be taken into consideration.

\section{Hamiltonian and current}

Let us remind briefly the properties of the graphene electron spectrum. The
tight-binding approximation allows to write down the following effective
Hamiltonian $2\times 2$ matrix
\begin{equation}
H(\mathbf{p})=\left(
\begin{array}{cc}
0 & h(\mathbf{p}) \\
h^{\ast }(\mathbf{p}) & 0%
\end{array}%
\right) ,  \label{ham}
\end{equation}%
where $h(\mathbf{p})=\gamma \left\{ e^{ip_{x}a}+2e^{-ip_{x}a/2}\cos {(p_{y}a%
\sqrt{3}/2)}\right\} $ and $\gamma =<\psi (a,0)|H_{0}|\psi (0,0)>$ is the $%
pp\pi $ transfer integral, which is of the order of $3$ eV. The dispersion
law of graphene, $\varepsilon _{1,2}(\mathbf{p})$ can be obtained by means
of its diagonalization:
\begin{eqnarray}
\varepsilon _{1,2}(\mathbf{p}) &=&\pm \gamma \left\{ 1+4\cos {(3p_{x}a/2)}%
\cos {(p_{y}a\sqrt{3}/2)}+\right.  \nonumber \\
&&\left. 4\cos ^{2}{(p_{y}a\sqrt{3}/2)}\right\} ^{1/2}.  \label{dra}
\end{eqnarray}%
In the vicinity of the $K$ point, the matrix elements of the effective
Hamiltonian can be expanded as
\[
h(\mathbf{p})=v_{0}(ip_{x}+p_{y}),
\]%
where $v_{0}=3\gamma a/2$. This results in the linear dispersion relation $%
\varepsilon _{1,2}(\mathbf{p})=\pm v_{0}p.$ Calculating the conductivity, we
can use this expansion for $\omega \ll \gamma \approx $ $3$ eV. Let us pass
to the calculation of the electric current, following the paper by Abrikosov
\cite{Ab}, where the dielectric function of Bi-type metals was considered. \
We assume that the external field is described by the vector potential $%
A_{j}.$ The current operator has the form
\begin{equation}
j_{i}(x)=e\tilde{\psi}^{+}(x^{\prime })v_{x^{\prime }x}^{i}\tilde{\psi}(x)-%
\frac{e^{2}}{c}\tilde{\psi}^{+}(x^{\prime })(m^{-1})_{x^{\prime }x}^{ij}%
\tilde{\psi}(x)A_{j}  \label{cur}
\end{equation}%
where $x^{\prime }\rightarrow x$, $v_{x^{\prime }x}^{i}$ and $%
(m^{-1})_{x^{\prime }x}^{ij}$ are the velocity and mass operators
correspondingly, the tilde in the notation ${\tilde{\psi}}$ means that the
operator is taken in the interaction representation with
\begin{equation}
V=-\frac{e}{c}\int \psi ^{+}(x^{\prime })v_{x^{\prime }x}^{i}\psi
(x)A_{i}(x)d^{4}x.  \label{inter}
\end{equation}

We will calculate the current in the linear approximation in $A_{i}(x)$. \
Therefore, the second term in Eq. (\ref{cur}) can be taken in zeroth
approximation with respect to the interaction (\ref{inter}). Expanding the
first term in Eq. (\ref{cur}) to the first order in $A_{i}(x)$, we get the
retarded correlator of four ${\tilde{\psi}}$ operators which has to be
averaged over the Gibbs ensemble. At finite temperatures, the Fourier
component (with respect to the spatial coordinates and imaginary time) of
this correlator%
\begin{equation}
\mathcal{P}\left( \omega _{l},\mathbf{k}\right) =\sum\limits_{\omega
_{n}}\int \frac{d^{2}\mathbf{p}}{\left( 2\pi \right) ^{2}}Tr\left\{ v^{i}%
\mathcal{G}\left( p_{+}\right) v^{j}\mathcal{G}\left( p_{-}\right) \right\}
\label{cor}
\end{equation}%
is expressed in terms of the temperature Green's functions:%
\[
\mathcal{G}\left( p\right) =\left[ i\omega _{n}-H(\mathbf{p})\right] ^{-1}
\]%
In Eq. (\ref{cor}) the notations $p_{\pm }$
are used, the summation is carried out over the fermionic frequencies $%
\omega _{n}=2\pi T\left( n+1/2\right) ,$ while the trace operation is
performed over the band index of the Hamiltonian $H(\mathbf{p})$. The latter
can be easily carried out in the representation where the Hamiltonian has
diagonal form:
\[
Tr\left\{ v^{i}\mathcal{G}v^{j}\mathcal{G}\right\} =v_{11}^{i}\mathcal{G}%
_{11}v_{11}^{j}\mathcal{G}_{11}+v_{22}^{i}\mathcal{G}_{22}v_{22}^{j}\mathcal{%
G}_{22}
\]%
\[
+v_{12}^{i}\mathcal{G}_{22}v_{21}^{j}\mathcal{G}_{11}+v_{21}^{i}\mathcal{G}%
_{11}v_{12}^{j}\mathcal{G}_{22}.
\]%
Then one can perform the summation over $\omega _{n}$ in standard way. For
instance, for the cross product of the Green functions one finds
\begin{eqnarray}
T\sum_{\omega _{n}}\mathcal{G}_{11}(p_{+})\mathcal{G}_{22}(p_{-}) =\frac{T}{%
i\omega _{l}-\varepsilon _{2}(\mathbf{p}_{+})+\varepsilon _{1}(\mathbf{p}%
_{-})} \\
\times \sum_{\omega _{n}}\left[ \frac{1}{i\omega _{-}-\varepsilon _{1}(%
\mathbf{p}_{-})}-\frac{1}{i\omega _{+}-\varepsilon _{2}(\mathbf{p}_{+})}%
\right]  \nonumber \\
=\frac{f_{0}[\varepsilon _{1}(\mathbf{p}_{-})]-f_{0}[\varepsilon _{2}(%
\mathbf{p}_{+})]}{i\omega _{l}-\varepsilon _{2}(\mathbf{p}_{+})+\varepsilon
_{1}(\mathbf{p}_{-})},  \nonumber  \label{sum}
\end{eqnarray}%
where we have taken into account that the photon frequencies $\omega
_{l}=2\pi lT$ are "even" in the Matsubara technique; $f_{0}$ is the Fermi
distribution function. Analytical continuation of the expressions similar to
Eq. (\ref{sum}) into the upper half plane of the complex frequency can be
performed by simple substitution $i\omega _{l}\rightarrow \omega +i\delta $
with $\delta \rightarrow 0$.

Let us notice that the current has to vanish when the vector potential does
not vary in time. Since the second term in Eq. (\ref{cur}) is time
independent, one can omit it, subtracting from the first term its value at $%
\omega =0$. In result%
\begin{eqnarray}  \label{con}
&&\sigma _{ij}(\omega ,k) = \frac{ie^{2}}{\pi ^{2}}\times \\
&& \left\{ \sum_{a=1,2}\int \frac{d^{2}pv^{i}v^{j}\{f_{0}[\varepsilon _{a}(%
\mathbf{p}_{-})]-f_{0}[\varepsilon _{a}(\mathbf{p}_{+})]\}}{[\varepsilon
_{a}(\mathbf{p}_{+})-\varepsilon _{a}(\mathbf{p}_{-})][\omega -\varepsilon
_{a}(\mathbf{p}_{+})+\varepsilon _{a}(\mathbf{p}_{-})]}\right.  \nonumber \\
&&\left. +2\omega \int \frac{d^{2}pv_{12}^{i}v_{21}^{j}\{f_{0}[\varepsilon
_{1}(\mathbf{p}_{-})]-f_{0}[\varepsilon _{2}(\mathbf{p}_{+})]\}}{%
[\varepsilon _{2}(\mathbf{p}_{+})-\varepsilon _{1}(\mathbf{p}_{-})]\{\omega
^{2}-[\varepsilon _{2}(\mathbf{p}_{+})-\varepsilon _{1}(\mathbf{p}%
_{-})]^{2}\}}\right\}.  \nonumber
\end{eqnarray}%
This expression acquired the factor 4 due to summation over spin and over
six points of the $K$ type (two per each Brillouin zone).

Hitherto we did not use any peculiarities of the graphene spectrum besides
the number of bands. Thus Eq. (\ref{con}) has a general character. For
graphene, the matrix of the velocity ${}$ near the point $K$ in the band
representation is determined by the Hamiltonian (\ref{ham}):
\begin{equation}
\mathbf{v}=v_{0}\left(
\begin{array}{cc}
0 & -\mathbf{e}_{y}+i\mathbf{e}_{x} \\
-\mathbf{e}_{y}-i\mathbf{e}_{x} & 0%
\end{array}%
\right) ,  \label{vel}
\end{equation}%
where $\mathbf{e}_{i}$ are unit vectors along the coordinate axis
directions. The unitary transformation which transforms the Hamiltonian from
the band representation to the diagonal one has the form
\[
U=\frac{1}{\sqrt{2}p}\left(
\begin{array}{cc}
p_{y}-ip_{x} & p_{y}-ip_{x} \\
-p & p%
\end{array}%
\right) .
\]%
Its application to the velocity matrix gives:
\[
U^{-1}\mathbf{v}U=\frac{v_{0}}{p}\left(
\begin{array}{cc}
\mathbf{e}_{x}p_{x}+\mathbf{e}_{y}p_{y} & i(\mathbf{e}_{x}p_{y}-\mathbf{e}%
_{y}p_{x}) \\
-i(\mathbf{e}_{x}p_{y}-\mathbf{e}_{y}p_{x}) & -\mathbf{e}_{x}p_{x}-\mathbf{e}%
_{y}p_{y}%
\end{array}%
\right) .
\]

The first term in Eq. (\ref{con}) corresponds to the intra-band
electron-photon scattering processes. In the limit of the high carriers
concentration $kv_{0}\ll (T,E_{F})$, it leads to the usual Drude expression
\begin{equation}
\sigma _{xx}^{intra}(\omega ,0)=\frac{e^{2}}{(i\omega -\tau ^{-1})}%
\sum_{a=1,2}\int \frac{d^{2}p}{2\pi ^{2}}v_{ax}^{2}\frac{df_{0}\left[
\varepsilon _{a}(\mathbf{p})\right] }{d\varepsilon },  \label{Drude}
\end{equation}%
with $1/\tau \rightarrow 0$ ($\tau $ is the transport scattering time). The
second term in Eq. (\ref{con}) owes its origin to the inter-band electron
transitions. The real part of this contribution (let us recall, that $%
i\omega _{l}\rightarrow \omega +i\delta $) at $k\rightarrow 0$ is reduced to
the well-known expression for the absorbed energy due to the direct
inter-band transitions.

\section{Asymptotic behavior of conductivity}

Let us pass to the discussion of the the pure graphene conductivity in
absence of gate voltage, when the chemical potential is equal to zero. The
integral (\ref{con}) can be performed analytically for various limiting
cases.

\subsection{Small spatial dispersion $kv_{0}\ll \protect\omega ,T$}

Putting $k=0$ and integrating over angle, one can find that the off-diagonal
elements of conductivity vanish, while the diagonal ones are equal to:
\begin{eqnarray}
\sigma _{xx}(\omega ,0) =\sigma _{yy}(\omega
,0)=-\frac{ie^{2}\omega }{\pi
}\times   \nonumber  \label{con1} \\
\left[ \frac{2}{\omega ^{2}}\int_{0}^{\infty }\varepsilon
d\varepsilon \left( \frac{df_{0}(\varepsilon )}{d\varepsilon
}\right) -\int_{0}^{\infty }d\varepsilon \
\frac{f_{0}(-\varepsilon )-f_{0}(\varepsilon )}{\omega
^{2}-4\varepsilon ^{2}}\right] .  \nonumber
\end{eqnarray}

The first (intra-band) contribution :
\begin{equation}
\sigma _{xx}^{intra}(\omega ,k)=2\ln 2\frac{ie^{2}T}{\pi \hbar \omega }%
,\quad kv_{0}<<\omega ,T  \label{con2}
\end{equation}%
(we write explicitly the Planck constant in the final expressions). This
result was obtained in Ref. \cite{W} using the Drude expression, Eq. ( \ref%
{Drude}).

The inter-band term in Eq. (\ref{con1}) has the form:
\begin{eqnarray}
\sigma _{xx}^{inter}(\omega ,0) &=&\frac{e^{2}}{4}\tanh \frac{\omega }{4T}
\label{con3} \\
&&+\frac{ie^{2}\omega }{\pi }P\int_{0}^{\infty }\frac{d\varepsilon }{\omega
^{2}-4\varepsilon ^{2}}\tanh \frac{\varepsilon }{2T},  \nonumber
\end{eqnarray}%
where P denotes the operation of taking the principal value of an integral.
This integral can be presented in more convenient form
\begin{eqnarray}
I_{P} &=&P\int_{0}^{\infty }\frac{d\varepsilon }{\omega ^{2}-4\varepsilon
^{2}}\left[ \tanh \frac{\varepsilon }{2T}-1\right]  \nonumber \\
&=&-2P\int_{0}^{\infty }\frac{d\varepsilon }{(\omega ^{2}-4\varepsilon
^{2})(\exp {(\varepsilon /T)}+1)}.  \nonumber
\end{eqnarray}

In the limit of low temperatures $T<<\omega $, the energies $\varepsilon
\sim T<<\omega $ are essential and we can take Taylor of the integrand over $%
T/\omega :$
\begin{equation}
I_{P}=-\frac{2T}{\omega ^{2}}[\ln 2+6\zeta (3)(T/\omega )^{2}],  \label{int1}
\end{equation}%
where $\zeta (3)=1.20$. One can see that the term with $\ln 2$
exactly cancels the intra-band contribution (\ref{con2}).

In the limit of high temperatures $T>>\omega $, the leading contribution to
the integral (\ref{con3}) originates from the region $\omega <\varepsilon <T$%
. One can obtain with the logarithmic accuracy:
\begin{equation}
I_{P}=-\frac{1}{8T}\int_{\omega }^{T}\frac{d\varepsilon }{\varepsilon }=-%
\frac{1}{8T}\ln {\frac{T}{\omega }}.  \label{con4}
\end{equation}

Collecting the Eqs. (\ref{con3}),(\ref{int1}), and (\ref{con4}), one can
write:
\begin{eqnarray}
\sigma _{xx}^{inter}(\omega,k) =\frac{e^{2}}{4\hbar }\tanh \frac{\omega }{4T}%
-\frac{2ie^{2}}{\pi \hbar }\times  \label{ht} \\
\left\{
\begin{array}{ll}
\displaystyle(T/\omega )[\ln {2}+6\zeta (3)(T/\omega )^{2}], &
kv_{0}<<T<<\omega , \\
\displaystyle(\omega /16T)\ln {(T/\omega )}, & kv_{0}<<\omega <<T.%
\end{array}%
\right.  \nonumber
\end{eqnarray}

\subsection{ Large spatial dispersion $\protect\omega \ll kv_{0},T$}

Now let us consider the limit of large dispersion $kv_{0}>>\omega
$. We choose the direction of $\mathbf{k}$ along the x-axis. In
the intra-band term of Eq. (\ref{con}), one can expand
\begin{eqnarray}  \label{itrlk}
\varepsilon _{2,1}(\mathbf{p}_{\pm }) &=&\sqrt{\varepsilon ^{2}\left(
p\right) +(v_{0}k/2)^{2}\pm pkv_{0}^{2}\cos \phi }\approx   \label{itrlk} \\
&&\varepsilon _{0}\left( p\right) \pm pkv_{0}^{2}\cos {\phi }/2\varepsilon
_{0},  \nonumber
\end{eqnarray}%
where $\varepsilon _{0}\left( p\right) =\sqrt{\varepsilon ^{2}\left(
p\right) +(v_{0}k/2)^{2}}$, $\varepsilon \left( p\right) =v_{0}p$, and $\phi
$ is the angle between $\mathbf{k}$ and $\mathbf{p}$.

Now we evaluate the integral in the Eq. (\ref{con}) for $\sigma
_{xx}^{intra}(\omega ,k)$. First of all, let us note that changing of the
variable of integration $\mathbf{p}\rightarrow -\mathbf{p}$ results in
\begin{eqnarray}
&&[\omega -\varepsilon _{a}(\mathbf{p}_{+})+\varepsilon _{a}(\mathbf{p}%
_{-})]^{-1}+[\omega +\varepsilon _{a}(\mathbf{p}_{+})-\varepsilon _{a}(%
\mathbf{p}_{-})]^{-1}  \nonumber \\
&=&\frac{2\omega }{\omega ^{2}-[\varepsilon _{a}(\mathbf{p}_{+})-\varepsilon
_{a}(\mathbf{p}_{-})]^{2}}.  \nonumber
\end{eqnarray}%
Using the expansion (\ref{itrlk}) one can see that in the significant domain
of integration, determined from the condition 
$\omega \ll pkv_{0}^{2}/\varepsilon _{0}\left( p\right) $, the
integrand of intra-band part $\sigma _{xx}^{intra}$ in Eq.
(\ref{con}), can be presented as follows
\[
v_{x}^{2}\frac{df_{0}(\varepsilon _{0})}{d\varepsilon _{0}}\frac{\omega
\varepsilon _{0}^{2}}{(pkv_{0}^{2}\cos {\phi })^{2}},
\]%
where $v_{x}=v_{0}\cos {\phi }$. Thus, $\cos {\phi }$ disappears
in the integrand. But for the intra-band contribution in the
transversal $\sigma _{yy}^{intra}$ component of conductivity, we
have $v_{y}=v_{0}\sin {\phi }$, and therefore, the longitudinal
and transversal components (respect with the $\mathbf{k}$ -
direction) are different:
\begin{eqnarray}
\sigma _{xx}^{intra}(\omega ,k) =\frac{-ie^{2}\omega }{\pi \hbar
T}\times
\nonumber  \label{trc} \\
\left\{
\begin{array}{ll}
\displaystyle(2T/kv_{0})^{2}\ln {2}, & \omega <kv_{0}<T, \\
\displaystyle\ln {(2}\sqrt{{kv_{0}T}}{/\omega )}\exp {(-kv_{0}/2T)}, &
\omega <T<kv_{0},%
\end{array}%
\right.   \nonumber
\end{eqnarray}

\begin{eqnarray}
\sigma _{yy}^{intra}(\omega ,k) =\frac{e^{2}}{\pi \hbar }\times  \nonumber
\label{trc} \\
\left\{
\begin{array}{ll}
\displaystyle\left( 4T/kv_{0}\right) \ln {2}, & \omega <kv_{0}<T, \\
\nonumber\displaystyle\sqrt{\pi kv_{0}/T}\exp {(-kv_{0}/2T)}, & \omega
<T<kv_{0}.%
\end{array}%
\right.  \nonumber
\end{eqnarray}

What concerns the inter-band contribution, here we can put
$\varepsilon
_{2,1}=\pm \varepsilon _{0}\left( p\right) $. Taking into account that $%
v_{12}^{x}=iv_{0}\sin \phi $ and integrating over $\phi $, one can obtain

\[
\sigma _{xx}^{inter}(\omega ,k)=\frac{-ie^{2}\omega }{4\pi }%
\int_{kv_{0}/2}^{\infty }\frac{d\varepsilon _{0}}{\varepsilon _{0}^{2}}\tanh
{\frac{\varepsilon _{0}}{2T}}.
\]%

Evaluating this integral, we obtain for the inter-band
contribution
\begin{eqnarray}
\sigma _{xx}^{inter}(\omega ,k) &=&\sigma _{yy}^{inter}(\omega ,k)=-\frac{%
ie^{2}\omega }{2\pi \hbar }\times   \nonumber  \label{interes} \\
&&\left\{
\begin{array}{ll}
\displaystyle(1/4T)\ln {(4T/kv_{0})}, & \quad \omega <kv_{0}<T, \\
\displaystyle1/kv_{0}, & \quad \omega <T<kv_{0}.%
\end{array}%
\right.   \nonumber
\end{eqnarray}

\section{Conclusions}

The expression Eq. (\ref{ht}) allows us to estimate the conductivity of pure
graphene, $\tau ^{-1}=0$. In this case, the first term plays the leading
role (for small $k$ and at low temperatures, when $\tanh \omega {/T}=1$:
\begin{equation}
\sigma (\omega ,k)=\frac{e^{2}}{4\hbar },\quad kv_{0}\ll T\ll \omega .
\label{uni}
\end{equation}%
Let us underline that this conductivity results from the electron
transitions between two intersecting bands at the $K$ points of the
Brillouin zone. Remarkable fact is that its value turns out to be universal,
independent of any parameter of graphene, like $\gamma _{0}$ or $v_{0}$.

Very recently the electric field effect in graphene was investigated
experimentally \cite{Novo}. Indeed, the universal conductivity behavior for
the samples with different carrier concentrations was found. In spite of
fact that the experiment was performed in conditions $\omega \ll T,$
different from Eq. (\ref{uni}), the minimal value of measured conductivity $%
\sigma ^{\left( \min \right) }=2e^{2}/\pi \hbar $ was found close to our
prediction.

At high temperatures, when the condition $kv_{0}\ll \omega \ll T$ is
fulfilled, the conductivity Eq. (\ref{con2}) becomes imaginary:
\[
\sigma (\omega ,k)=2i\ln 2\frac{e^{2}T}{\pi \hbar \omega }
\]%
and it depends on temperature.

The behavior of such type can be observed on experiments involving
the plasmon modes. The dispersion law of such modes for 2D systems
is gapless,
\[
\omega =v_{0}\sqrt{\kappa k},\quad \kappa =\frac{2e^{2}T\ln {2}}{\hbar
^{2}v_{0}^{2}\varepsilon _{\infty }},
\]%
where $\varepsilon _{\infty }$ is the lattice dielectric constant.

So far we considered the 2D graphene sheet. The results obtained can be
immediately applied to the 3D graphite if one neglect the interaction
between the layers. Then the integration with respect to the $p_{z}$
component of the quasi-momentum in the Brillouin zone gives just the
additional factor $1/c_{z}$ in comparison with the conductivity of graphene,
where $c_{z}$ is the distance between the layers in the $z$-direction. For
instance, Eqs. (\ref{Drude}) and (\ref{con2}) acquire the factor $1/c_{z}$.
In the low-frequency limit $\omega \ll 1/\tau $, conductivity can be
estimated as
\[
\sigma ^{\left( 3\right) }(\omega ,k\rightarrow 0)=2\ln 2\frac{e^{2}T\tau }{%
\pi \hbar c_{z}}.
\]%
The number of phonons in the 2D graphene at low temperatures ($T\ll T_{D},$ $%
T_{D}$ is the Debye temperature) is proportional to $T^{2}.$ Since the Fermi
surface is assumed to be small ($\varepsilon _{F}<T<T_{D}$), all scattering
angles are essential in this region of temperatures. For the electron
collision rate, which is determined by the electron-phonon interaction, one
can write $\tau ^{-1}=\alpha T^{2}/T_{D}$ with the constant $\alpha $ of the
order of unity. Thus the in-layer resistivity turns out to be linear in
temperature:
\begin{equation}
\rho =\frac{\pi \hbar \alpha c_{z}T}{2e^{2}T_{D}\ln {2}}=320\,\frac{\alpha T%
}{T_{D}}\,\mu \Omega \,\mathrm{cm}.  \label{grres}
\end{equation}%
According to Ref. \cite{ESM}, $\rho $ = 60 $\mu \Omega $\thinspace cm at $%
T=300$ K in agreement with the above estimate ($T_{D}\simeq 2000K$ for
graphite, $\alpha \sim 1$). The equation (\ref{grres}) answers the question
discussed in Refs. \cite{ESM,Ar} on the pressure dependence of the graphite
resistivity. We see that the resistivity decreases under the pressure
because the inter-layer distance $c_{z}$ decreases and the Debye temperature
$T_{D}$ grows.

\begin{acknowledgments}
This work was supported by the RFBR, Grant No 04-02-17087. \ The authors are
grateful to D.V. Livanov for valuable discussions.
\end{acknowledgments}

\end{document}